\documentclass[aps, pre, twocolumn, floats, showpacs, longbibliography]{revtex4-1}

\usepackage{graphicx}
\usepackage{hyperref}
\usepackage{amsmath}

\usepackage{xcolor}

\begin{document}
	\title{The impact of heterogeneity on the co-evolution of cooperation and epidemic spreading in complex networks}
	
	\author{Mehran Noori$^{1}$, Nahid Azimi-Tafreshi$^{1}\footnote{Corresponding: nahid.azimi@iasbs.ac.ir}$, and Mohammad Salahshour$^{2,3,4}$}
	\affiliation{$^{1}$Department of Physics, Institute for Advanced Studies in Basic Sciences (IASBS), Zanjan 45137-66731, Iran}
	\affiliation{$^{2}$Department of Collective Behavior, Max Planck Institute of Animal Behavior, Konstanz, Germany}
	\affiliation{$^{3}$Center for the Advanced Study of Collective Behaviour, University of Konstanz, Konstanz, Germany}
	\affiliation{$^{4}$Department of Biology, University of Konstanz, Konstanz, Germany}
	
	\date{\today}
	
	\begin{abstract}

		The dynamics of herd immunity depend crucially on the interaction between collective social behavior and disease transmission, but the role of heterogeneity in this context frequently remains unclear. Here, we dissect this co-evolutionary feedback by coupling a public goods game with an epidemic model on complex networks, including multiplex and real-world networks. Our results reveals a dichotomy in how heterogeneity shapes outcomes. We demonstrate that structural heterogeneity in social networks acts as a powerful catalyst for cooperation and disease suppression. This emergent effect is driven by highly connected hubs who, facing amplified personal risk, adopt protective strategies out of self-interest. In contrast, heterogeneity in individual infection costs proves detrimental, undermining cooperation and amplifying the epidemic. This creates a ``weakest link'' problem, where individuals with low perceived risk act as persistent free-riders and disease reservoirs, degrading the collective response. Our findings establish that heterogeneity is a double-edged sword: its impact is determined by whether it creates an asymmetry of influence (leverage points) or an asymmetry of motivation (weakest links), recommending disease intervention policies that facilitate cooperative transition in hubs (strengthening the leverage point) and homogenize incentives to weakest links.

	\end{abstract}
	\maketitle
	\section{Introduction}

The study of the dynamical evolution of diseases within a population and their control methods has attracted significant attention in the social sciences \cite{Anderson, Keeling, modeling}. Protective measures, which often entail individual costs, can effectively curb the spread of diseases and help achieve herd immunity \cite{Steinegger, Fredrik, Zhen, gosak2021community}. Evolutionary game theory provides a framework to model individual decisions regarding these protective measures \cite{Newton, tanimoto2015wp}. A growing body of research focuses on how evolutionary game theory and epidemic dynamics interact, recognizing that personal choices to reduce disease risk can be viewed as strategic decisions under infection risks \cite{chang2020ug, huang2022qh, amaral2021, wang2020uc, wang2015coupled, wang2020vaccination, khanjanianpak2022ap}. These coevolutionary models frequently reveal emergent phenomena, such as abrupt transitions or the emergence of oscillations in both infection density and protection levels, driven by the feedback loop between behavioral and epidemiological processes.

Although these protective measures impose private burdens, they also endow collective protection against disease and benefit society as a whole. Hence, the achieved herd immunity can be interpreted as a public good, and therefore the public goods game (PGG) can be used to analyze how and under what conditions populations evolve toward higher levels of cooperation that reduce overall infection risk \cite{dees2018, yong2021, morison2024}. In this regard, we recently introduced a co-evolutionary dynamics by coupling the public goods game and the susceptible-infected-susceptible (SIS) epidemic model in a well-mixed population and also on a regular lattice structure \cite{prevwork}. Our results showed that the presence of an epidemic can lead to the emergence of cooperation in regimes where cooperation does not naturally evolve in the standard PGG. In turn, higher cooperation can positively impact the control of the epidemic dynamics. While these results shed light on the co-evolutionary dynamics of cooperation and epidemics in an idealized, homogeneous world, real-world interactions show different sources of heterogeneity. Motivated by this observation, here, we consider two sources of heterogeneity: structural heterogeneity in social networks and heterogeneity in the infection cost.

Complex networks provide the most effective description of the heterogeneous structures found in real-world systems \cite{barabasi, BOCCALETTI, newman, moez2006}. Epidemic models on highly heterogeneous networks often show lowered or even vanishing epidemic thresholds. This means that disease can spread and persist even when the infection rate is very low, as highly connected hubs facilitate transmission  \cite{pastorsisnet, pastor2015, wang2024}. Similarly, the evolution of cooperation on complex networks indeed shows that heterogeneity promotes cooperation by enabling cooperative behavior to survive and even thrive under conditions unfavorable in more homogeneous settings \cite{Santos2008}. Additionally, in a more complex scenario, individuals may cooperate in groups that do not necessarily risk direct disease transmission. For example, infection spreading might occur on a physical contact layer, while cooperation dynamics may take place on a different, social network layer. Such structures can be modeled using multiplex networks, where a set of nodes can be connected through two types of links, or layers \cite{multilayer, multiplex}. This separation allows the study of how disease dynamics and cooperation co-evolve or influence each other without being confined to the same network structure.

Similarly, in the real world, the cost of infection is not the same for everyone and varies due to many factors. Variations in infection burden can arise, for instance, from behavioral differences, contact structure and environmental exposures, individual economic status, and disease severity, which affect both risk of infection and the magnitude of personal costs \cite{infectioncost1, infectioncost2, infectioncost3}.

These considerations raise the important question that how different sources of heterogeneity affect the co-evolutionary dynamics of epidemics and cooperation. To address this question, here, we begin by investigating the co-evolutionary dynamics of a PGG and epidemic spreading on complex networks, comparing the dynamics on regular, random, and scale-free structures. We demonstrate that structural heterogeneity significantly improves cooperation and disease suppression. We provide an explanation for this phenomenon by showing hubs act as strategic leverage points, and explore the model on multiplex and real-world networks to test the robustness of our findings. Finally, we investigate heterogeneity in the cost of infection, revealing that it has a detrimental effect on cooperation. We provide an explanation for this phenomena based on the ``weakest link'' problem. Our work provides an understanding of how different forms of heterogeneity shape socio-epidemiological outcomes.

	\section{\label{sec2}Model}
	We consider the model introduced in \cite{prevwork}. Here, we focus on the dynamics of the model on complex networks. A population of $N$ individuals engages in a PGG while simultaneously experiencing an SIS epidemic. A susceptible individual (S) becomes infected by its infected neighbors with a baseline probability $\alpha_{0}$, and an infected (I) individual recovers with probability $\mu$ \cite{pastor2015}. Individuals are also either cooperators (C) or defectors (D). Cooperators contribute a cost $c_{G}$ to a public good, which is multiplied by an enhancement factor $r$ and distributed equally among all group members \cite{PGG}. This defines four possible states for an individual $i$: $s_{i}\in\{SC,SD,IC,ID\}$.
	
	To couple the game and epidemic dynamics, we assume cooperation reduces the risk of infection. Two parameters capture this: the self-interested preventive impact $\alpha_{r}$, which reduces the infection probability for susceptible cooperators, and the altruistic preventive impact $\alpha_{t}$, which reduces the infectiousness of infected cooperators. Thus, the infection probability varies based on the strategies of both interacting individuals. The probability that a susceptible cooperator gets infected when in contact with an infected defector is reduced to $\alpha_0\alpha_r$. Similarly, a defector gets infected with probability $\alpha_0\alpha_t$ when in contact with an infected cooperator and with probability $\alpha_0$ when in contact with an infected defector. When both infectious and susceptible agents are cooperators, the disease spreads with probability $\alpha_0\alpha_r\alpha_t$.
	
	Infected individuals also incur a direct infection cost, $c_I$. This cost represents treatment or other burdens. Therefore, the payoffs in a group of $g$ individuals with $N_c$ cooperators (excluding the focal individual) are:
	\begin{equation}
		\begin{split}
			\pi_{IC}^g&=rc_G\frac{N_c+1}{G}-c_G-c_I,\\
			\pi_{ID}^g&=rc_G\frac{N_c}{G}-c_I,\\
			\pi_{SC}^g&=rc_G\frac{N_c+1}{G}-c_G,\\
			\pi_{SD}^g&=rc_G\frac{N_c}{G}.
		\end{split}
		\label{payoff}
	\end{equation}
Here, $G$ is the size of the group.

Evolutionary and epidemic updates occur on different timescales, governed by a parameter $\tau$. With probability $\tau$, an individual updates their strategy; with probability $1-\tau$, their disease state is updated. For a strategy update, a focal individual $x$ participates in $k+1$ public goods games, each centered around one of its neighbors or itself, where $k$ is the degree (number of neighbors) of the focal node $x$. The payoff, $\Pi_x$, is determined by summing up the gains from all the groups in which player $x$ is involved. Next, another individual $y$ is randomly selected from $x$'s neighbors, and its payoff, $\Pi_y$, is calculated in the same manner. The focal individual $x$ then updates its strategy $s_x$ by adopting the strategy $s_y$ of node $y$ based on the Fermi probability, given as follows:
	\begin{equation}
		W(s_x\leftarrow s_y)=\frac{1}{1+[\exp(\Pi_x-\Pi_y)/K]},
		\label{fermi}
	\end{equation}
	where, $K$ controls the level of noise. 
	
	If with probability $1-\tau$, the selected node $x$ is chosen to participate in the epidemic dynamics, it will recover with probability $\mu$ if it is currently infected. Alternatively, it may become infected by one of the infected neighbors in the network with probability $\alpha_0$, $\alpha_0\alpha_r$, $\alpha_0\alpha_t$, or $\alpha_0\alpha_r\alpha_t$, depending on its strategy and the strategy of its neighbor.
	
	Repeating these elementary steps for $N$ times constitutes one full Monte Carlo step (MCS), which ensures that every node has an opportunity to update once on average. The Monte Carlo steps continue until the dynamics reach a stationary state.
	\begin{figure}[t!]
		\centering
		\includegraphics[width=\linewidth]{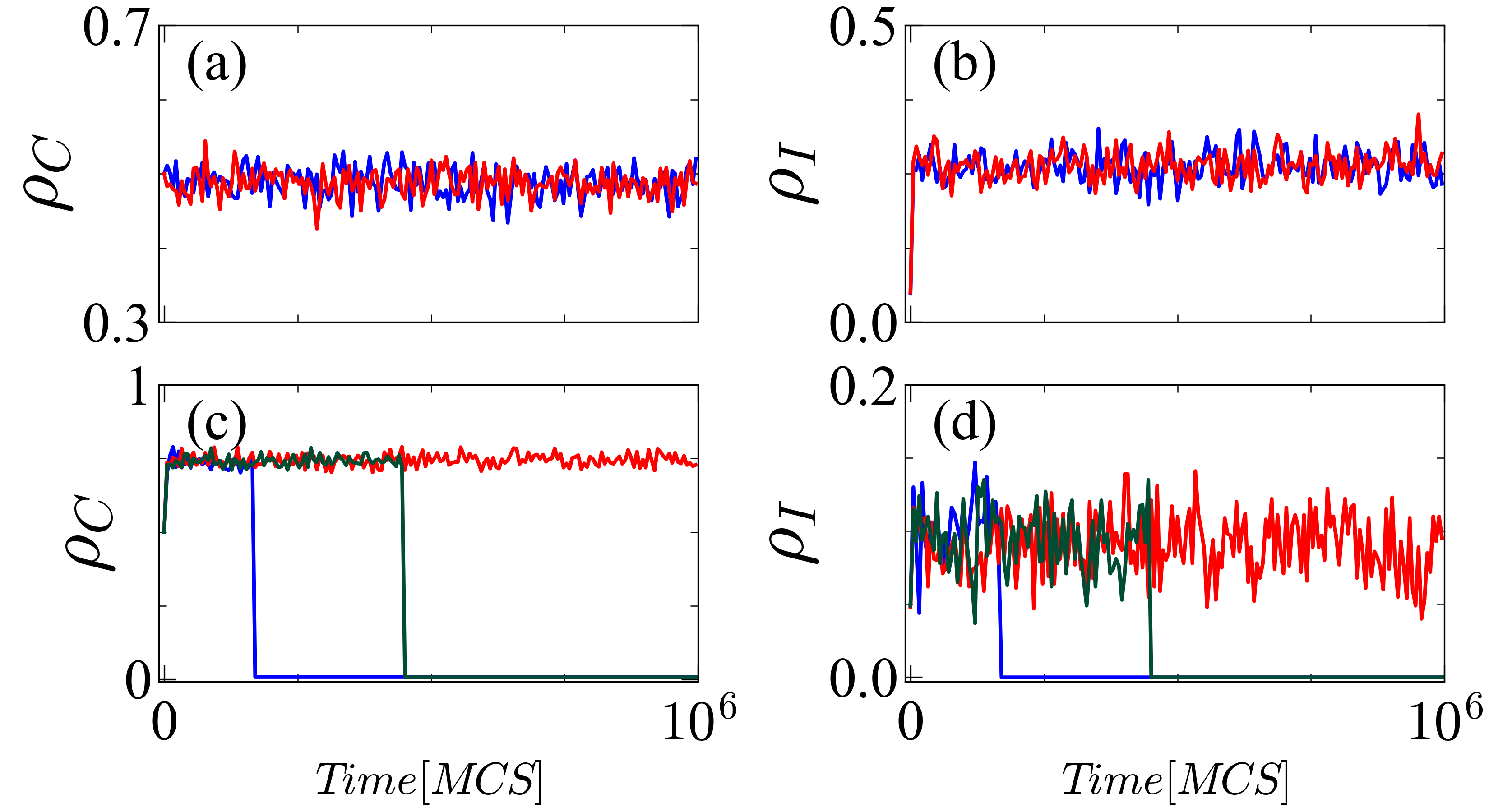}
		\caption{Time evolution of the fraction of cooperators $\rho_C$ and infected individuals $\rho_I$ on a random network with average degree $\langle k \rangle = 4$, over some realizations. The enhancement factor is $r = 1.5$ in (a)-(b) and $r = 2.6$ in (c)-(d). For the larger $r$, some of the realizations fall into the absorbing state with vanishing densities. Model parameters are set to $\alpha_0 = 0.5$, $\alpha_r = \alpha_t = 0.01$, $c_I = 10$, $c_G = 1$ and $\tau = 0.01$. }
		\label{fig1}
	\end{figure}
	   \begin{figure*}[t]
		\centering
		\includegraphics[width=\linewidth]{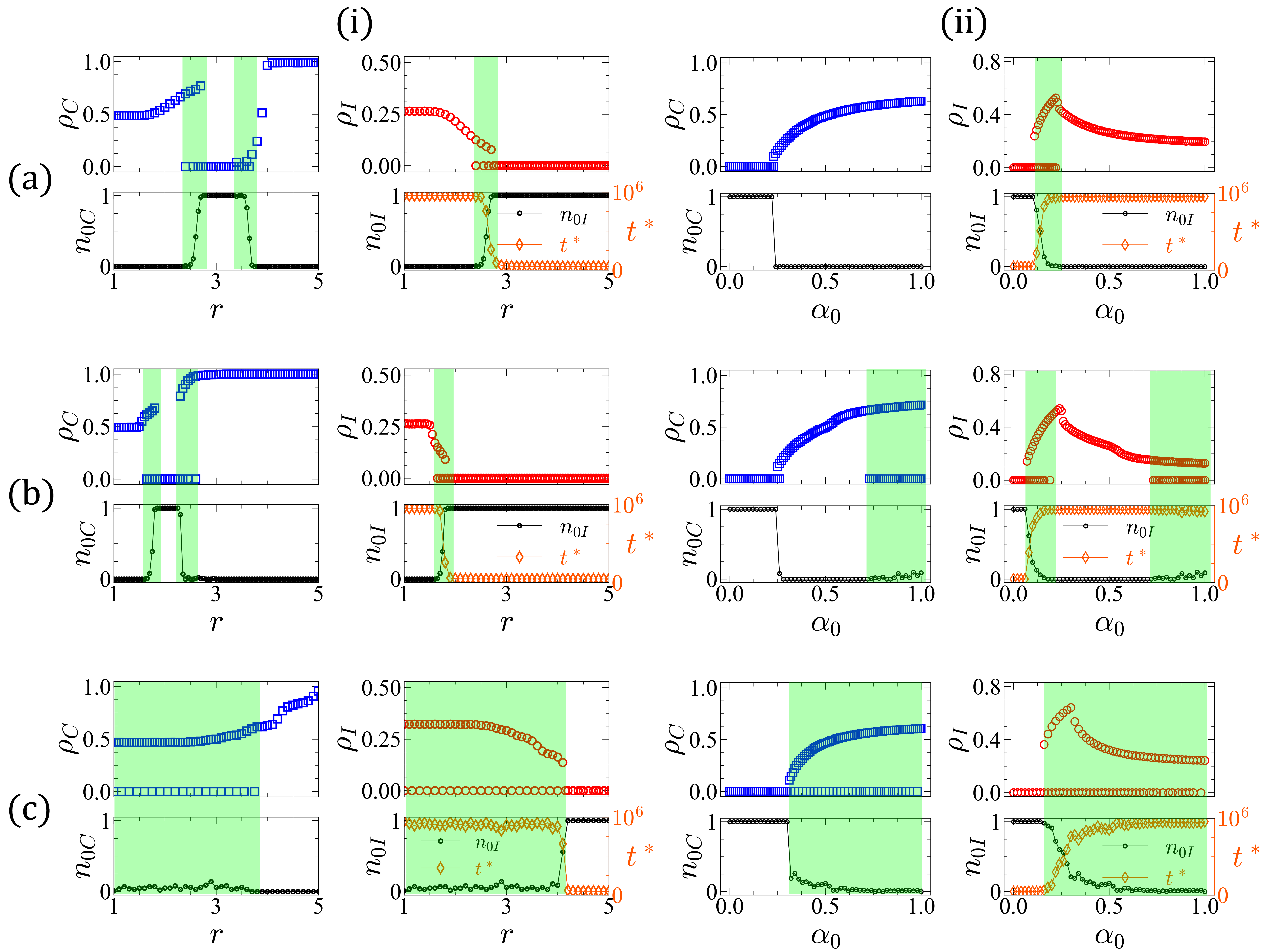}
		\caption{The coevolutionary dynamics of cooperation and epidemics on the ER random network (a), BA network (b), and SQ lattice (c). The stationary values for the fraction of cooperators and infected individuals as functions of the enhancement factor, $r$, with a fixed disease transmission probability $\alpha_0 = 0.5$ (i), and the infection transmission probability $\alpha_0$, with a fixed enhancement factor $r=1.5$ (ii), are plotted. Bottom panels show $n_{0C}$ and $n_{0I}$, the fraction of simulations where the zero steady-state solutions (absorbing states) are reached. In the bottom right panel, we also plot the average disease lifetime of epidemics, $t^*$. The green-shaded regions denote the bistable region. Across all networks, increasing the enhancement factor reduces disease spreading. However, the epidemics are eradicated for a smaller enhancement factor on more heterogeneous networks, indicating the beneficial effect of heterogeneity for curbing epidemics by cooperation. Similarly, increasing the disease transmission rate, $\alpha_0$, reduces infection due to the evolution of higher cooperation. Results are averaged over 100 simulations, each simulated for $10^6$ time steps, with the time averages taken over the last $10^3$ time steps. The ER and BA networks are of size $N=1000$ with the average degree of $ \langle k \rangle = 4$. The size of the SQ lattice is $N=1024$, with Von Neumann connectivity.
	}
		\label{fig2}
	\end{figure*}
	\section{\label{sec3} The Role of Structural Heterogeneity}
	\subsection{single complex networks}
	
We begin by considering the model on Erd\H{o}s--R\'enyi (ER) random network and the Barab\'asi–Albert (BA) scale-free network and analyze the model's dynamics in comparison with results obtained on a regular square lattice (SQ). The size of the square lattice is $N=32^2=1024$, while both the random and scale-free networks consist of $N = 1000$ nodes.
	
\textbf{Phase diagram.} We present the time evolution of cooperation and density of infected individuals on a random ER network with average degree $\langle k \rangle = 4$ in Fig. \ref{fig1}. We choose two different enhancement factors, both below the level at which cooperation would normally appear in the standard public goods game. When the enhancement factor is small, cooperation arises from the interplay between cooperative behavior and epidemic dynamics, resulting in non-zero densities within the system. However, for the larger values of $r$, we find that in some realizations the system falls into the absorbing state where the densities vanish. In other words,  as $r$ increases, bistability emerges, where two distinct outcomes can be observed. This phenomenon is also observed in scale-free networks.

In Fig. \ref{fig2}, we summarize the model's behavior, showing how the densities of cooperators and infected individuals vary across different network structures. In Fig.~\ref{fig2}((a),(i)) we present the stationary values of cooperation, $\rho_C$, and the fraction of infected individuals, $\rho_I$, as a function of the enhancement factor $r$ in an ER network. The parameter $\alpha_0$ is set in such a way that the spread of disease in the standard SIS model is significant. Hence, cooperation emerges at small values of $r$ due to the presence of the disease. We call this cooperative state a non-trivial cooperative state, since it does not typically emerge at a standard PGG. As $r$ increases, the bistability emerges. The green-shaded regions indicate bistability, where the absorbing and non-zero density states coexist. $n_{0C}$ and $n_{0I}$ represent the fraction of realizations in which the densities of cooperators and infected individuals, respectively, fall to zero. Outside the green-shaded area, the values of $n_{0C}$ and $n_{0I}$ are zero or one, indicating that only one of the steady-state solutions exists. However, in the overlapping regions, they change continuously between the two states. We note that, by increasing the network size, the number of realizations approaching zero in the bistable region decreases and the width of the bistable regions shrinks (see Fig.~S1 in the Supplementary).
	
Here, we also present the average disease lifetime $t^*$ which gradually decreases in the bistable region, leading to the emergence of the disease-free state. After the epidemic subsides, cooperation is no longer necessary, and the cooperative population in the random networks approaches zero when the enhancement factor $r$ is below the threshold value of the standard PGG in this network. However, as the parameter $r$ increases further, cooperation re-emerges at the threshold of the standard public goods game. Here we observe another bistable region in the population of cooperators, and the absorbing states appear in a small region of $r$. In the scale-free structure (Fig.~\ref{fig2} ((b),(i))), the bistable regions are present as well. Nevertheless, the transition point for cooperation from a non-trivial state to zero, as well as the transition point for the resurgence of cooperation to its trivial state, occurs at smaller values of $r$. This observation suggests that the level of non-trivial cooperation and epidemic prevalence is closely related to the structural heterogeneity of the network. As we can see in Fig.~\ref{fig2} ((c),(i)), both nonzero and absorbing states exist within the non-trivial regime in the SQ lattice. Also, the amount of cooperation in this network is often not enough to eradicate the disease in the non-trivial region. By contrast, heterogeneity not only leads to higher cooperation and stronger suppression of epidemics, but it also leads to a monostable region where epidemics are fully eradicated.

In Fig.~\ref{fig2}((a)-(c),(ii)), the densities are presented in terms of infection probability $\alpha_0$. For small values of $\alpha_0$, the disease fails to spread, and consequently, no cooperation evolves. As $\alpha_0$ increases, we observe a bistable region such that the epidemic and disease-free states coexist. However, once the epidemic starts to spread, individuals can improve their payoffs through cooperation. Consequently, with increasing $\alpha_0$, cooperation evolves at a later phase transition. Surprisingly, further increasing $\alpha_0$ beyond this second transition point leads to a reduction in the fraction of infected individuals. Also, as it is known \cite{pastorsisnet}, the transition point to the epidemic state occurs earlier on networks with heterogeneous structures. Therefore, cooperative behavior starts at a smaller value of $\alpha_0$ and higher overall levels of cooperation occur as network heterogeneity increases. Furthermore, at high values of $\alpha_{0}$, disease-free states appear in some realizations, both in the scale-free and square lattice, although the number of these states is not significant. To better visualize the impact of coupling the two dynamics, in the Supplementary Information (Fig.~S2), we compare the results with scenarios where the standard PGG and the SIS epidemic spreading operate independently as single dynamics on the same networks.

	\begin{figure}[t]
		\centering
		\includegraphics[width=\linewidth]{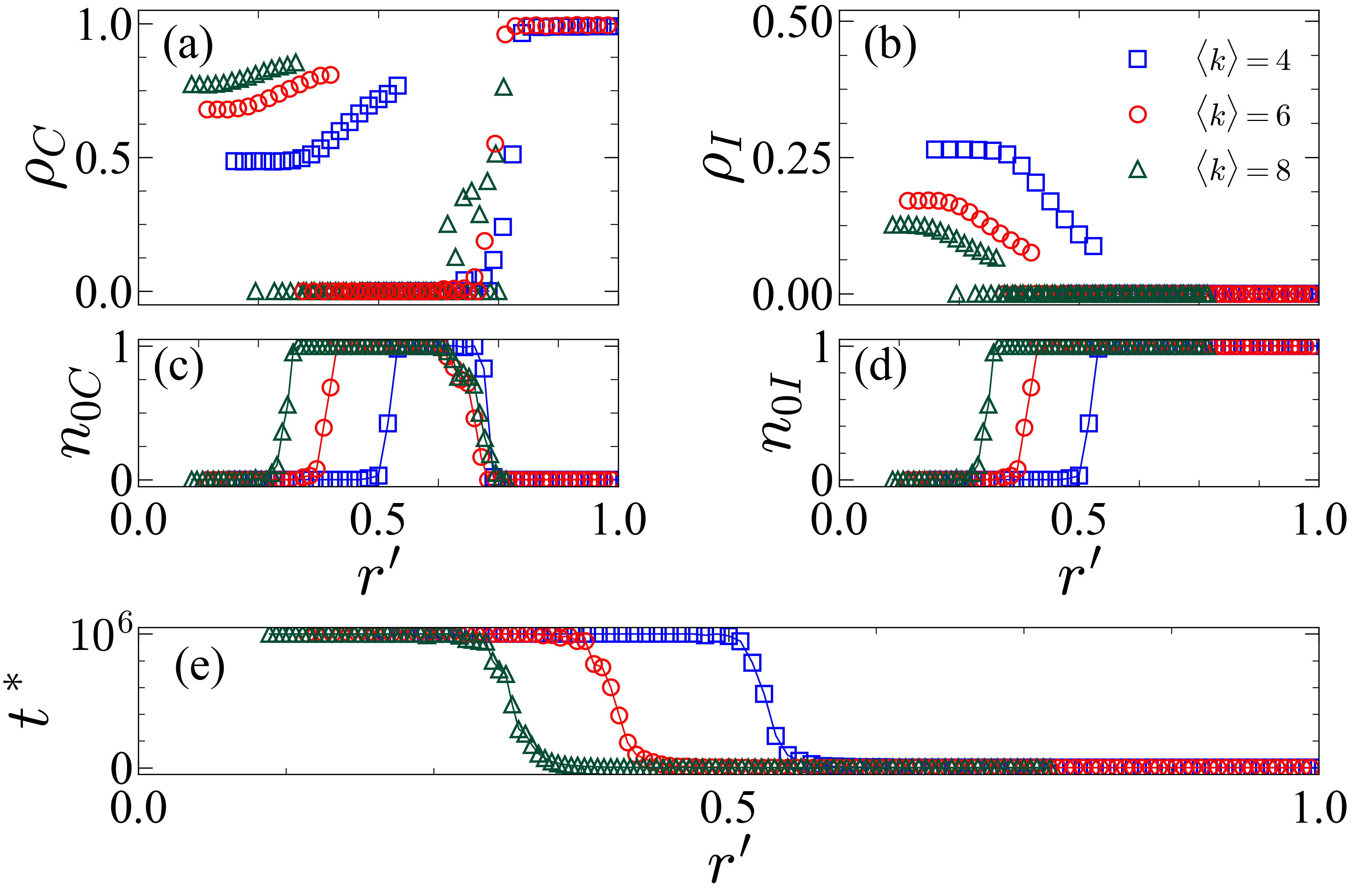}
		\caption{The fraction of cooperators $\rho_C$ (a), infected individuals $\rho_I$ (b), absorbing states for the density of cooperators $n_{0C}$ and infected individuals $n_{0I}$, respectively (c)-(d) and the average disease lifetime $t^*$ (e) as functions of the normalized enhancement factor $r'$ in random ER network, for three different average degrees. Increasing the mean network degree ($\langle k \rangle$) promotes cooperation by amplifying the risk of the epidemics. This increases the health incentive for individuals to cooperate, driving the system toward disease eradication earlier than in sparse networks. Model parameters are $\alpha_0=0.5$, $\alpha_r = \alpha_t = 0.01$, $c_I=10$, $c_G=1$, and $\tau=0.01$. The network size is $N = 1000$. Results are averaged across 100 realizations, each run for $10^6$ time steps, with time averages calculated over the final $10^3$ steps. }
		\label{fig3}
	\end{figure}	
	
\begin{figure}[t]
		\centering
		\includegraphics[width=\linewidth]{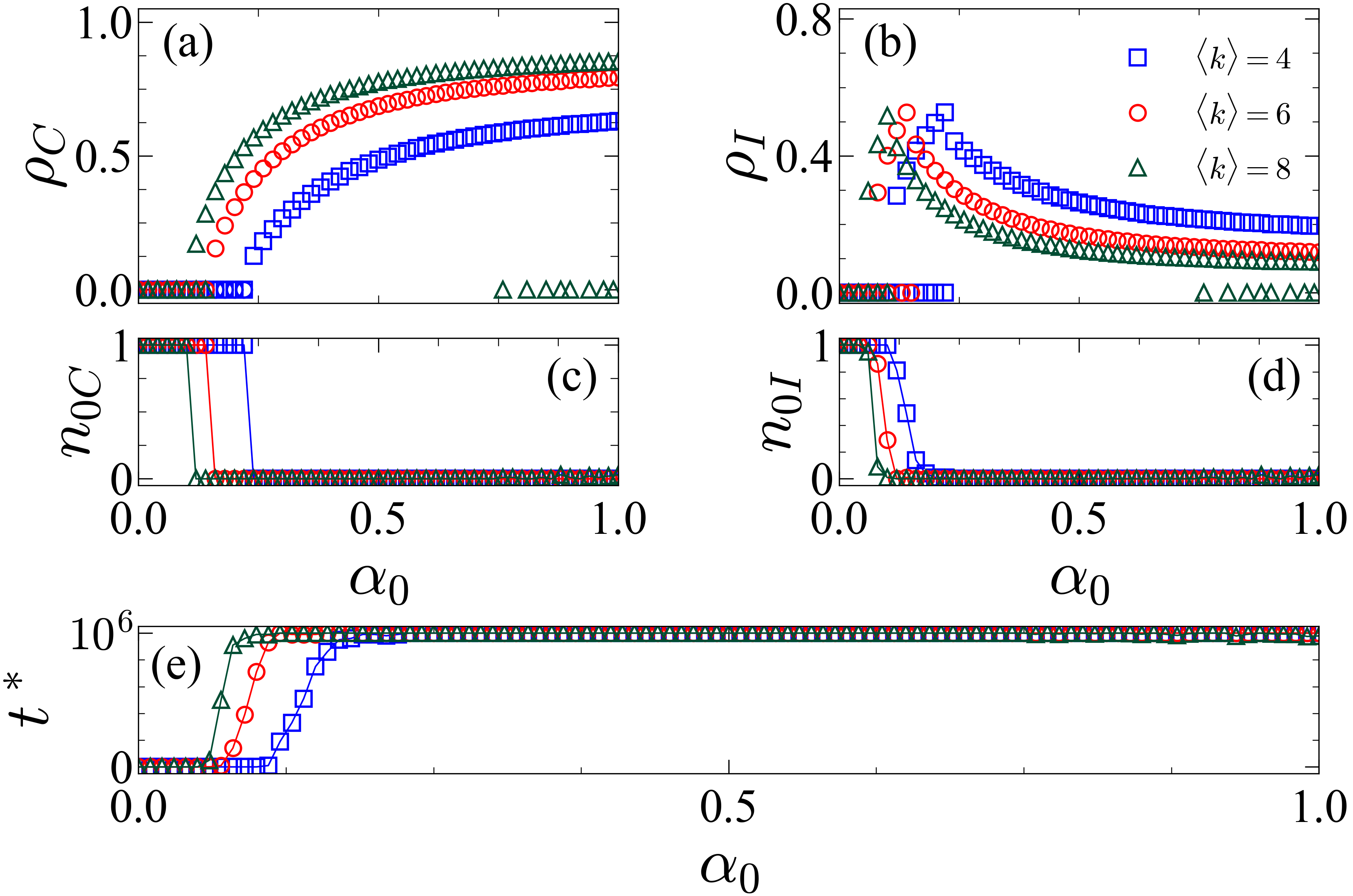}
		\caption{The fraction of cooperators $\rho_C$ (a), infected individuals $\rho_I$ (b), absorbing states for the density of cooperators $n_{0C}$ and infected individuals $n_{0I}$, respectively (c)-(d) and the average disease lifetime $t^*$ (e) as functions of disease spreading probability $\alpha_0$ in random ER network, for three different average degrees. As the baseline probability of infection ($\alpha_0$) rises, the population responds with a surge in cooperation. This adaptive behavioral response acts as a social immune system, effectively curbing the epidemic prevalence even as the biological transmissibility of the disease increases. Model parameters are $r=1.5$, $\alpha_r = \alpha_t = 0.01$, $c_I=10$, $c_G=1$, and $\tau=0.01$. The network size is $N = 1000$. Results are averaged across 100 realizations, each run for $10^6$ time steps, with time averages calculated over the final $10^3$ steps.}
		\label{fig4}
	\end{figure}

\begin{figure}[t]
		\centering
		\includegraphics[width=\linewidth]{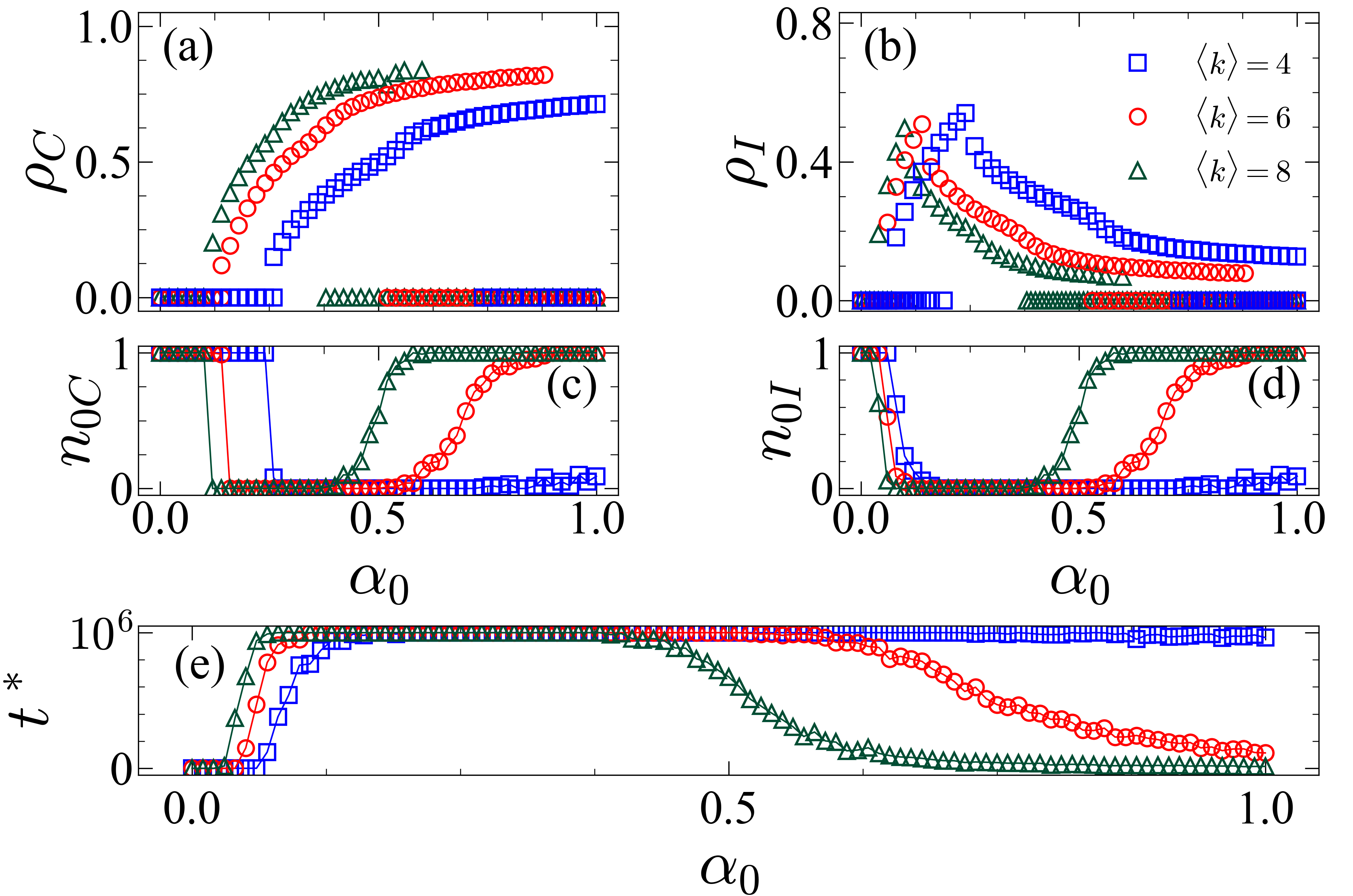}
		\caption{The fraction of cooperators $\rho_C$ (a), and infected individuals $\rho_I$ (b), absorbing states for the density of cooperators $n_{0C}$ and infected individuals $n_{0I}$, respectively (c)-(d) and the average disease lifetime $t^*$ (e) as functions of disease spreading probability $\alpha_0$ in BA scale-free network, for three different average degrees. In scale-free networks, highly connected nodes face the greatest risk and thus lead the charge in cooperation. Increasing connectivity in heterogeneous networks leads to a precipitous drop in infection, as hubs transform from super-spreaders into super-blockers. Model parameters are $r=1.5$, $\alpha_r = \alpha_t = 0.01$, $c_I=10$, $c_G=1$, and $\tau=0.01$. The network size is $N = 1000$. Results are averaged across 100 realizations, each run for $10^6$ time steps, with time averages calculated over the final $10^3$ steps.}
		\label{fig5}
	\end{figure}

\textbf{The effect of mean degree.}	In Figs.~\ref{fig3}, \ref{fig4}, and \ref{fig5}, we examine the impact of the mean degree of networks on the dynamics of the model. By changing the network connectivity, the public goods group size changes, which leads to changes in the threshold value of the enhancement factor above which cooperation is no longer a social dilemma ($r_{threshold}=g$). To keep our results across networks comparable, we thus consider the normalized enhancement factor $r'=\frac{r}{\left\langle k \right\rangle +1 }$ as the control parameter. Figure \ref{fig3}(a)-(b) presents the phase transitions in the densities of cooperative and infected individuals as functions of the normalized enhancement factor for an ER network. As we observe, an increase in the average degree enhances cooperation, leading to a further reduction in disease and resulting in a transition point to the disease-free state at lower $r'$ values. Additionally, while the width of the bistability region shows little sensitivity to the average degree, the bistability regime emerges at lower values of $r'$ as the average degree increases. The bistability region is observed through the behavior of the fraction of absorbing states $n_{0C}$ and $n_{0I}$ (Fig.~\ref{fig3}(c)-(d)). Furthermore, the average disease lifetime $t^*$ is shorter in structures with a higher average degree (Fig.~\ref{fig3}(e)), due the fact that higher connectivity leads to a faster equilibration of the dynamics. The same behaviour is observed in BA scale-free networks (Fig.~S3 in the Supplementary). However, these networks demonstrate sharper transitions that occur at lower $r'$ values compared to those seen in random ER networks.
	
	
Similarly, in Fig.~\ref{fig4} we present the behavior of the system on ER networks in terms of disease transmission probability $\alpha_0$, with increasing mean degree. In networks with higher average degrees, cooperation is strengthened, resulting in a notable reduction in epidemic levels, even at high probabilities of disease transmission. This contrasts with the standard SIS disease model, where an increase in the average degree typically leads to a higher prevalence of the disease as the probability of transmission rises. However, as observed from the behavior of the disease's lifetime $t^*$, in random networks with a high infection transmission probability, the disease persists in the network, and even high levels of cooperation cannot completely eradicate it.
		
In BA scale-free networks, a similar behavior is observed for population densities, such that the onset of cooperation occurs at lower $\alpha_0$ values with increasing mean degree, leading to phase transitions at smaller probabilities and consequently better control of disease prevalence (Fig.~\ref{fig5}). Additionally, as seen from the behavior of $n_{0C}$ and $n_{0I}$, the bistability region reappears at higher disease transmission values with increasing average degree. Interestingly, in scale-free networks with high mean degree, the emergent cooperation can eradicate the disease at high values of $\alpha_0$. Specifically, as the average degree increases, the disease's lifetime $t^*$ drops to zero at high values of $\alpha_0$. This means that despite a strong likelihood of transmission, the disease eventually vanishes from the system.
	
The beneficial effect of mean connectivity on the evolution of cooperation and suppression of epidemics may be surprising, as this phenomenon does not occur when each dynamics is considered in isolation. When epidemics dynamics is considered in isolation, higher connectivity increases the risk of desease transmission, ultimately leading to stronger spreading. Similarly, higher connectivity, by leading to stronger mixing, favors defection in public goods dynamics in the absence of epidemics. However, when the two are coupled, the exact opposite in both dynamics is observed due to an efficient feedback loop between the two dynamics: higher connectivity increases the transmission risk, providing a higher advantage for cooperation. The observation that higher connectivity fosters greater cooperation and mitigates epidemic severity stems from the fact that the coupled dynamics are ultimately driven by the efficacy of cooperation in disease suppression.
	
\textbf{The leverage point.} Our results can be explained by an analytical treatment of heterogeneity. The earlier onset of the epidemic in heterogeneous networks compared to more homogeneous ones results from the fact that for the SIS model on an uncorrelated network, the epidemic threshold is defined by the basic reproduction number $R_0 = (\alpha_0 / \mu) \cdot (\langle k^2 \rangle / \langle k \rangle) > 1$ \cite{pastorsisnet, pastor2015}. For scale-free networks with a degree distribution exponent $2 < \gamma \le 3$, the second moment of the degree distribution, $\langle k^2 \rangle$, diverges with system size, leading to a vanishing epidemic threshold in the thermodynamic limit \cite{barabasi}. This makes such networks exceptionally vulnerable to disease invasion. Paradoxically, the very structure that facilitates disease spread also contains the seeds of its suppression, when epidemic dynamics is coupled with the cooperation dynamics. The high degree of hubs amplifies their exposure to infection, making the personal cost of infection, $c_I$, a dominant term in their payoff calculation. This creates a powerful, self-interested incentive for these highly connected individuals to adopt the protective cooperator strategy. This selfish choice, however, has a profound collective benefit. When hubs cooperate, they benefit from both reduced susceptibility ($\alpha_r$) and reduced transmissibility ($\alpha_t$), effectively acting as targeted immunization agents at the most critical nodes of the network. This ``emergent altruism" dramatically lowers the system's effective $R_0$, explaining the potent disease suppression and even eradication observed in BA networks (Fig.~\ref{fig5}(e). Structural heterogeneity thus creates strategic leverage points, where the amplified self-interest of a few produces a disproportionate public good.

	\subsection{Multiplex Structures}
	\begin{figure}[t]
		\centering
		\includegraphics[width=\linewidth]{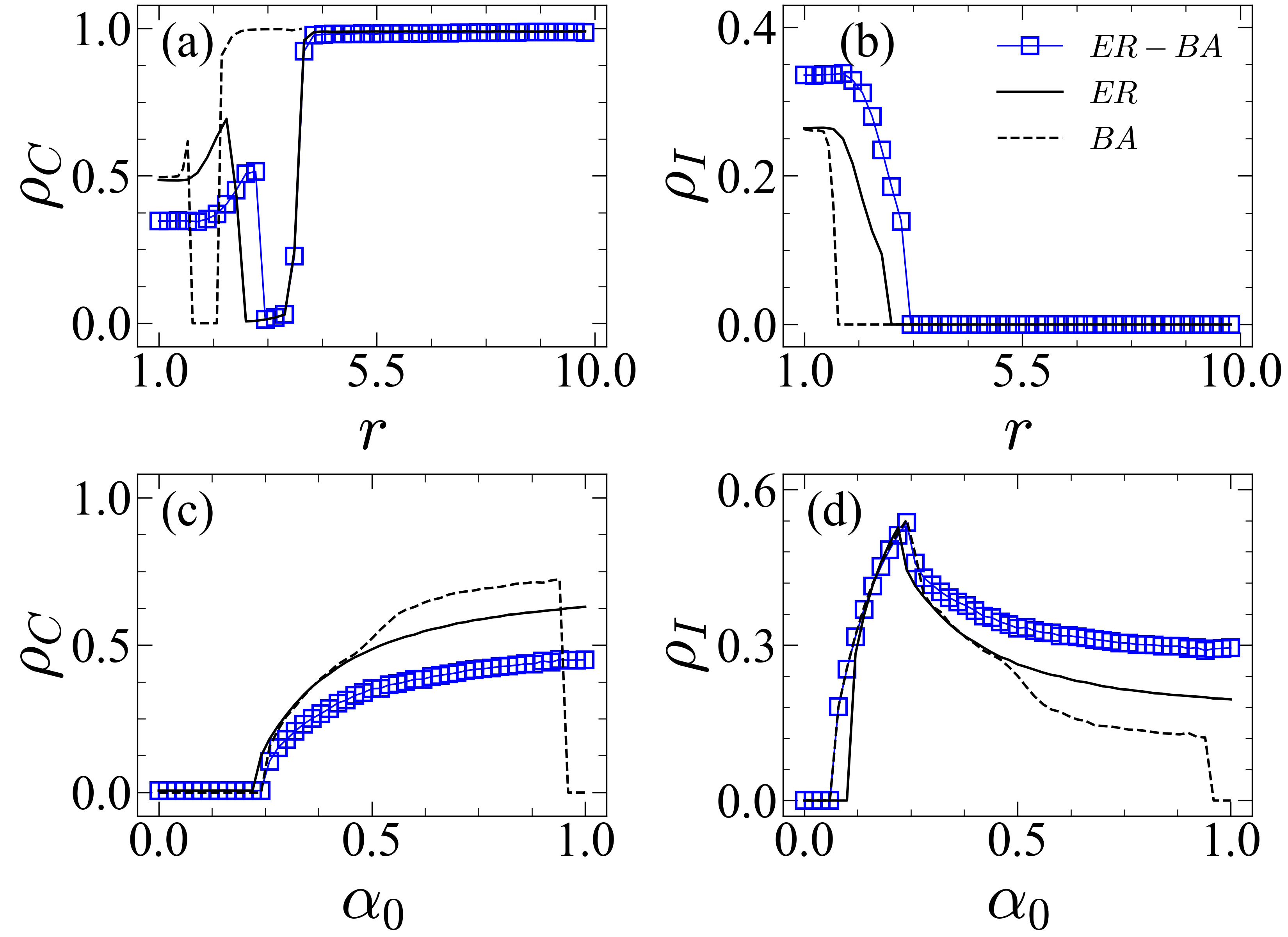}
		\caption{Comparison of the densities of cooperators $\rho_C$ and infected individuals $\rho_I$ on a multiplex network versus single-layer networks as functions of the enhancement factor $r$, with $\alpha_0=0.5$ (a)-(b) and as functions of $\alpha_0$ with $r = 1.5$ (c)-(d). In the multiplex network, the game dynamics take place on a random ER network, and the epidemic spreads on a BA scale-free network. The model’s behavior on the multiplex network (blue squares) is compared to that on a single-layer random network (solid line) and a single-layer scale-free network (dashed line). The mean degree is fixed at $\left\langle k\right\rangle = 4$ in all cases. This comparison reveals that separating the layers of social influence and physical contact (multiplex) weakens the collective response. }
		\label{fig6}
	\end{figure}
\begin{figure*}[t]
	\centering
	\includegraphics[width=\linewidth]{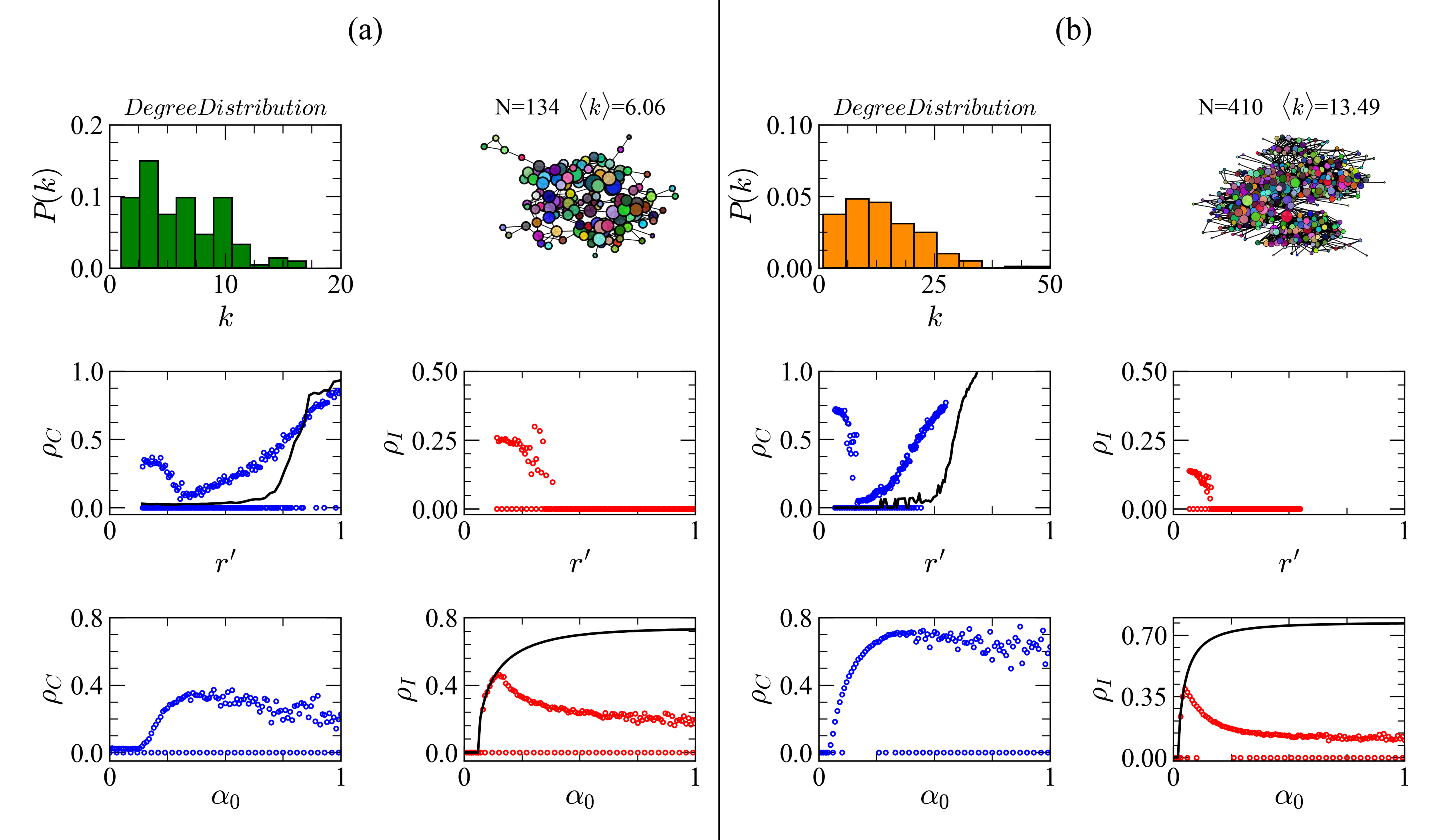}
	\caption{The model dynamics on real-world networks. The degree distribution and the behavior of density of cooperators and infected individuals are presented for friendships interactions in a French high school with size $N = 134$  and mean degree $\left\langle k \right\rangle = 6.06$ (a) and social interactions during the infectious at science gallery in Dublin with $N = 410$ and $\left\langle k \right\rangle = 13.49$ (b). The black curves show the cooperator density as a function of $r'$ in a single public goods dynamic, and the fraction of infected individuals as a function of $\alpha_{0}$ in a single SIS epidemic dynamic on these networks. The real-world contact patterns exhibit the same feedback loops and phase transitions predicted by the theoretical models, showcasing the framework's applicability to actual human social dynamics.}
	\label{fig7}
\end{figure*}
	
	In many realistic scenarios, the game dynamics and epidemic spreading can take place on different networks. In other words, social networks that influence strategic choices and those driving disease transmission may be distinct entities. This can occur, for instance, when the nature of disease spreading or cooperative interactions is limited to different types of interaction (e.g., physical versus virtual networks). 
	
	In the following, we consider the model on a multiplex network in which a set of nodes is connected with two types of links. By structuring the model this way, one layer of the multiplex network represents physical contacts that facilitate disease spread, while the other layer reflects social influences on behavioral decisions. To thoroughly explore a variety of outcomes, we have constructed a scenario using two networks with different structures. We begin by assuming that the epidemic spreads on the BA scale-free network, while the process of cooperation plays out on the ER random network.
	
	In Fig.~\ref{fig6}, the results of the model on a multiplex network are compared with the scenario where the model operates on a single random or scale-free network. For visibility, the absorbing states in the bistable region are not shown. In the non-trivial region, the emergence of cooperation is reduced, leading to a higher prevalence of disease when the model's dynamics are on the multiplex network. This is because the individuals who are most influential in the game dynamics are not necessarily the hubs of disease transmission, decoupling the strategic incentives from the epidemiological risks. Notably, a similar trend is observed when the network roles are reversed. That is, when the game takes place on a scale-free network, and the epidemic spreads on a random network (Fig.~S4 in the Supplementary). Even when two networks share the same underlying structure, whether random or scale-free with identical degree distributions, similar results can be expected. This is because the nodes in the two layers do not necessarily have the same degree, and the local neighborhoods of the nodes differ.
	
	\subsection{\label{sec4} Real-World Networks}
	
	We complement our analysis of the role of structural heterogeneity by applying the model to empirical networks derived from human contact patterns. Empirical networks reflect real interaction patterns, which frequently exhibit community clustering, the presence of super-spreaders, and evolving contact sequences \cite{friends,dublin, dublin2, dublin3}. These elements are essential for understanding both the spread of the epidemic and social behavior changes. Here we consider two real networks: 
		\begin{itemize}
		\item first network is based on self-reported friendship relations collected in a French high school in Marseille \cite{friends}. In this dataset, students indicate their social ties through a survey, resulting in a static friendship network.
		
		\item second network used was generated from a scientific experiment in which participants were equipped with wearable sensors capable of detecting close-range face-to-face proximity. The sensors continuously recorded contact events between individuals, and these interactions were aggregated to construct time-resolved networks representing the dynamics of human social contacts \cite{dublin, dublin2, dublin3}.
		\end{itemize}
			
    These two networks show different types of human interactions: one with stable social connections in a school, and the other with short, spontaneous contacts in a public space. This gives complementary views for studying how processes, like the spread of disease, move through social networks. 

	Figure \ref{fig7} illustrates the degree distribution and the dynamics of the model for these real-world networks. The model's behavior on these structures is qualitatively consistent with previous results, such as the existence of different phases and the beneficial effect of mean network connectivity (compare the two networks with different mean connectivity). In the non-trivial region (small $r'$), the presence of disease enhances cooperation, and a decline in disease prevalence at high values of $\alpha_0$ is observed due to the emergence of cooperation. As observed, in the coupled dynamics, the population of cooperating individuals is enhanced compared to the single public goods dynamics, leading to a significant reduction in disease prevalence compared to the single disease spread dynamics. In the coupled dynamics on these networks, the small size allows for the observation of absorbing states in the region where non-zero solutions for the densities of cooperators and infected individuals are present.

	
	\section{\label{sec5}Heterogeneous Infection Cost}
	\begin{figure}[t]
		\centering
		\includegraphics[width=\linewidth]{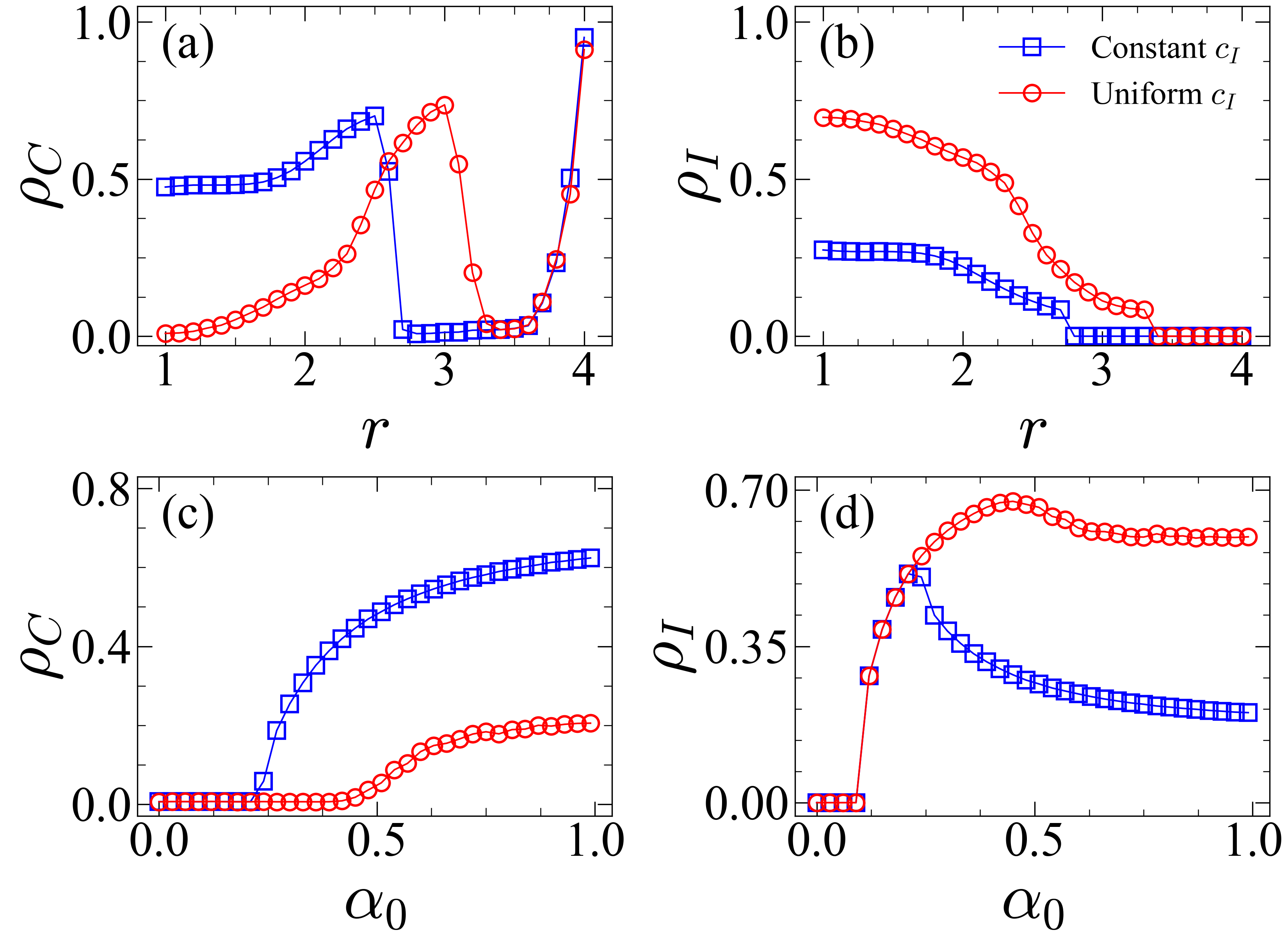}
		\caption{The density of cooperators $\rho_C$ and infected individuals $\rho_I$ as functions of enhancement factor $r$ (a)-(b) and disease spreading probability $\alpha_0$ (c)-(d) in ER random network with $N = 1000$ and $\langle k \rangle = 4$. The curves are plotted for the constant infection cost $c_I=5$ and the uniformly randomly distributed $c_I \in [0, 10]$. When costs are uniform, cooperation is robust; however, when costs vary, low-risk individuals act as free-riding disease reservoirs (weakest links), dragging down the collective cooperation level and allowing the epidemic to persist. Model parameters are $r = 1.5$, $\alpha_0=0.5$, $\alpha_r = \alpha_t = 0.01$, $c_G = 1$, and $\tau = 0.01$. Results are averaged over 100 realizations, each simulated for $10^5$ time steps, with the time averages taken over the last 1000 time steps.}
		\label{fig8}
	\end{figure}

	So far, we have analyzed the impact of heterogeneity in the structure of networks on the dynamics of the model. Here, we intend to explore heterogeneity in the cost of infection, which we have considered constant up to this point. However, in the real world, different individuals do not necessarily incur the same costs when they become infected. To investigate the influence of this heterogeneity, we consider $c_I$ values are chosen from a uniform random distribution in the interval $[0,10]$.

	Fig. ~\ref{fig8} presents the results of this modification on the model's dynamics within an ER random network. Here we compare the dynamics resulting from heterogeneous infection costs with the baseline case where $c_I=5$. In Fig.~\ref{fig8}(a)-(b), which displays the behavior of the model as a function of the enhancement factor $r$, we observe a notable decrease in the cooperation level when $c_I$ is uniformly randomly distributed. As we can observe, a greater enhancement factor is necessary for the complete elimination of the disease and then achieving full disease control results in cooperators density tending towards zero at larger enhancement factor values. Similarly, as Fig.~\ref{fig8}(c)-(d) demonstrates, cooperation emerges at higher values of infection probability and is less effective in controlling epidemics at higher values of $\alpha_0$. Similar results are observed when studying this type of heterogeneity in BA and SQ networks (see Figs.~S5 and S6 in the Supplementary Information).
	
	This phenomenon can be understood as a ``weakest link'' problem, a concept from public goods theory where the collective outcome is determined by the minimum individual effort \cite{weaklink}. Individuals with a low personal $c_I$ have little incentive to pay the fixed cost of cooperation, $c_G$. They become persistent defectors, acting as free-riders and, crucially, as reservoirs for the pathogen. Their behavior increases the overall infection risk for everyone, lowering the average payoff for cooperators and destabilizing the cooperative state. Thus, while structural heterogeneity is leveraged by its most influential members (hubs), cost heterogeneity causes the system to be dragged down by its least motivated members.

	\section{Discussion}
	\label{conclusion}
	
In this work, we investigated the effect of structural and cost heterogeneity on the coevolutionary dynamics of the SIS epidemic model and the public goods game. We examined the model on both single and multiplex complex networks, as well as some real-world structures. Our results reveal a fundamental dichotomy in how heterogeneity affects the coevolutionary dynamics of cooperation and epidemics. 
	
Our finding indicates that cooperation is higher in random and scale-free networks compared to regular networks. This is because random and scale-free networks have more heterogeneous contact patterns, which facilitate cooperation. Consequently, disease spread is reduced in these networks at a lower value of the enhancement factor. In scale-free networks, due to the presence of high-degree hubs, as the enhancement factor grows large, the cooperative efforts become overwhelmingly effective, leading to the abrupt disappearance of the disease, and cooperation suddenly stops. However, in multiplex networks, where strategic and epidemiological interactions are separated into distinct layers, the decoupling of these processes typically suppresses cooperation, triggering higher infection spread.
	
Beyond structural network heterogeneity, infection cost heterogeneity adds another critical layer that hampers the effectiveness of cooperation in suppressing epidemic outbreaks. This disorder in costs weakens collective cooperative responses because those with lower perceived costs may cooperate less or selfishly, reducing the overall level of cooperation needed to effectively curb the spread.

Our work provides clear policy recommendations. Our ``Leverage Point'' analysis indicates that hubs have the highest intrinsic motivation to cooperate. Therefore, facilitating their transition to cooperation (e.g., prioritized vaccination) yields disproportionate systemic benefits. On the other hand, to counter the ``Weakest Link" effect, policies should aim to homogenize the perceived cost of infection. This could involve targeted education for low-risk groups (increasing perceived $c_I$) or subsidies for protection costs ($c_G$) to ensure even the weakest links find it rational to cooperate. Furthermore, the multiplex analysis suggests that interventions are most effective when the social influence network overlaps with the physical contact network. Community-based interventions (where neighbors influence neighbors) are superior to broad, impersonal mandates that decouple social pressure from physical risk.
		
In summary, individuals within a system exhibit diverse behaviors and connections, and this heterogeneity significantly influences the dynamics and outcomes of coupled social and disease spreading processes. Recognizing these differences is crucial for accurately modeling and understanding how epidemics unfold within complex social networks. These insights can guide the development of policies that effectively manage epidemic risk by promoting and leveraging cooperative behaviors that help reduce disease transmission and enhance collective health outcomes.

\section{Acknowledgement.} M.S. acknowledges funding from German Research Foundation (DFG – Deutsche Forschungsgemeinschaft)
under Germany’s Excellence Strategy - EXC 2117-422037984.



	\bibliographystyle{unsrt}
	\bibliography{refs}
\end{document}